# Auto Tuning of Hadoop and Spark parameters

Mrs. Tanuja Patanshetti[1], Mr. Ashish Anil Pawar[2], Ms. Disha Patel[3], Mr. Sanket Thakare[4]

[1]*Assistant Professor, College of Engineering, Pune, Wellesley Rd, Shivajinagar, Pune, Maharashtra 411005*

[1] trp.comp@coep.ac.in, [2] pawaraa16.it@coep.ac.in, [3] patelds16.it@coep.ac.in

***Abstract*** — *Data of the order of terabytes, petabytes, or beyond is known as Big Data. This data cannot be processed using the traditional database software, and hence there comes the need for Big Data Platforms. By combining the capabilities and features of various big data applications and utilities, Big Data Platforms form a single solution. It is a platform that helps to develop, deploy and manage the big data environment. Hadoop and Spark are the two open-source Big Data Platforms provided by Apache. Both these platforms have many configurational parameters, which can have unforeseen effects on the execution time, accuracy, etc. Manual tuning of these parameters can be tiresome, and hence automatic ways should be needed to tune them.*

*After studying and analyzing various previous works in automating the tuning of these parameters, this paper proposes two algorithms - Grid Search with Finer Tuning and Controlled Random Search. The performance indicator studied in this paper is Execution Time. These algorithms help to tune the parameters automatically. Experimental results have shown a reduction in execution time of about 70% and 50% for Hadoop and 81.19% and 77.77% for Spark by Grid Search with Finer Tuning and Controlled Random Search, respectively.*

## I. INTRODUCTION

Nowadays, lots of data is being generated in terabytes and petabytes. The traditional data management platform couldn’t manage these large amounts of data, and hence the Big Data Platforms came into the picture. Big Data Platforms could easily process and analyze large and complex data. The quality that caused the Big Data Platforms to dominate the traditional data management platforms was that they distributed data processing tasks across multiple computers. It also helped in reducing the cost of storage of the data. Apache Hadoop [1] and Apache Spark [2] are the two big data platforms worked upon in this paper. Apache Hadoop and Apache Spark both have many parameters to be tuned before executing any job. Various approaches like Machine Learning [9], Container-based solutions [10], using different block sizes [11], etc., have been proposed earlier to tune the parameters of the Hadoop. While the solutions are based on approaches like Neural Network [12], Simulated Annealing Algorithms [13], Artificial Neural Network, Support Vector Regression and Decision Tree [14], Machine Learning [6], etc. have been proposed for self-tuning of Apache Spark. Methods such as using locality-sensitive hashing algorithm [15] have also been adopted to improve the performance of Hadoop in the past.

Apache Hadoop [1] is an open-source big data platform used widely among various organizations due to its enormous processing power. It has clusters of the commodity hardware on which it runs the applications. Looking over the importance of Hadoop, optimizing its performance becomes a key issue to focus on. Hadoop has about 200 parameters that require tuning before any job is run on it. Tuning these parameters requires expert knowledge, and even many experts find it difficult to tune these parameters. Also, it is seen that not all the parameters have a significant impact on the performance of Hadoop. There are only a few which impact the system performance. Tuning these few parameters in the right fashion will improve the system's accuracy, throughput, and execution time. All these parameters need tuning whenever a new job is submitted. So, manual tuning these parameters can be a very difficult task. Hence, a method to auto-tune these parameters is required.

Apache Spark [2] is also an open-source big data platform that stores, manages, processes big data in a distributed way. Spark is considered to be developer-friendly given that it can be worked with different languages like Python, Java, Scala, and R. It can be used for batch processing, stream processing, etc. Like Hadoop, Spark also has around 180 parameters that need to be tuned for better system performance. Expert knowledge is required to tune these parameters as well, and once they are tuned to their best values, it can improve various performance indicators. Only very few parameters affect the system performance, and manual tuning these parameters is a tedious task. Thus, autotuning these parameters has become a necessity.

In this paper, two algorithms have been proposed that help in auto-tuning the parameters. The algorithms are Grid Search with Finer Tuning and Controlled Random Search. The system also includes a Configuration Manager and Performance Evaluator (CMPE) module, which connects the system and the algorithms modules. The paper identifies 12 parameters of Hadoop and 11 parameters of Spark, which significantly affect system performance. Grid Search with finer tuning sends the values of the parameters to CMPE by sampling, then CMPE adjusts these values in the system and then runs Hadoop/Spark. After getting the results, it returns the execution time and the configuration of the parameter used. Controlled Random Search also sends the values of the parameters to the CMPE in a random manner. The CMPE then works as mentioned above. Simultaneously, a log file is also maintained, which helps keep track of various execution times and the respective configuration used. With the help of a log file, tracing any error also becomes easier. After analyzing the log file, the best possible configuration was observed, and then running the job on that configuration significantly reduced the execution time.
The paper is organized as follows. Section II and III deal with how the parameters were chosen based on the previous work. Section IV and V deal with the cluster setup and system design. Initial experiments were performed, and their results are mentioned in section VI. CMPE, the two algorithms, the experiment methodology, and their results are described in sections VII, VIII, and IX. Section X explains the comparison of two algorithms with each other and also with Hadoop and Spark. Section XI and XII deals with the conclusion and future scope. And the last section, XIII, shows the references used.

## II. PREVIOUS WORK

The parameters which are used in this paper are those which have a significant effect on the performance of the system. 12 parameters for Hadoop and 11 parameters for Spark have been selected based on the research done previously in the Big Data Laboratory at the College of Engineering, Pune. In this research, various feature selection algorithms were used to identify the parameters that affect the system's performance. The algorithm used in this paper was - Ensemble Feature Selection using Chained Gaussian Processes and Gradient Boosted Decision Tree Regression.
It gave the important features as the output. After finding the parameters that affected the system's performance, they were tuned manually (Manual tuning 1 parameter whilst keeping others constant), and the results were analyzed. It was proved in this research that these sets of parameters (mentioned in Section III of this paper) were bound to improve the system performance by up to 30%.

## III. PARAMETERS

Configuration Parameters are settings that affect Hadoop and Spark behavior which in turn affect its performance. Configuration Parameters have default values and explicitly run on these default values when not set to any value. On default parameters, the performance might not be optimum as it depends on various factors.

### *A. Hadoop*

Users of Hadoop can set around 200 parameters, but all the parameters don't impact the performance at all or have less impact. So, after studying these parameters' performance impact, the paper shortlists 12 parameters to use in the proposed approaches.

1. Mapreduce.map.memory.mb:
   The amount of memory requested by each map task from the scheduler.
2. dfs.blocksize(Bytes):
   The size of the blocks used for new files.
3. Mapreduce.tasktracker.map.tasks.maximum:
   The task tracker simultaneously runs a certain number of map tasks. This parameter configures the maximum number of these.
4. mapreduce.job.reduce.slowstart.completedmaps:
   Expressed as a fraction of number map tasks to be completed before starting the reduce phase.
5. Mapreduce.map.output.compress:
   Decides whether to compress output from the map phase before sending it across the network.
6. mapreduce. Job. Reduces
   Decides the number of reduced tasks in a job
7. mapreduce.task.io.sort.mb
   It is expressed in Megabytes as the amount of memory used by the buffer when file sorting is done.
8. mapreduce. Job. maps:
   Decides the number of map tasks in a job.
9. mapreduce.task.io.sort.factor:
   In the sorting process, this parameter decides the number of streams to merge in one go. Also, it is the number of open files.
10. dfs.replication:
    The actual number of replications when the file is generated.
11. mapreduce. Task tracker.reduce.tasks. Maximum:
    A maximum number of reduce tasks that run simultaneously by a task tracker.
12. Mapreduce.job.jvm.numtasks:
    Decides the number of tasks being run per JVM. If specified as -1, there is no limit.

Following Table-I presents these parameters with their range and default value.

**Table I**

| Parameter Name | Default Value | Range |
|---|---|---|
| mapreduce.map.memory.mb | 1024 | [256,3072] |
| dfs.blocksize(Bytes) | 128 | [32, 256] |
| Mapreduce.tasktracker.map. tasks.maximum | 2 | [2,128] |
| Mapreduce.job.reduce. slowstart.completedmaps | 0.05 | [0.025,0.9] |
| Mapreduce.map. output.compress | FALSE | [FALSE,TRUE] |
| mapreduce.job.reduces | 1 | [1,4] |
| mapreduce.task.io.sort.mb | 100 | [32,128] |
| mapreduce.job.maps | 2 | [2,32] |
| mapreduce.task.io.sort.factor | 10 | [5,80] |
| dfs.replication | 3 | [1,3] |
| mapreduce.tasktracker.reduce.tasks.maximum | 2 | [2,128] |
| mapreduce.job.jvm.numtasks | 1 | [1,1024] |

***B. Spark***

Configuration Parameters of Spark control application settings. Spark has over 150 configurable parameters, out of which 11 are shortlisted that have a significant impact on performance.

1. Spark.task.cpus:
   The number of CPU cores to be allocated for each task
2. spark. memory.storageFraction:
   Fraction of Java Heap Space as storage space.
3. Spark.network.timeout (seconds):
   All interactions in the network follow the timeout set by this parameter.
4. Spark.memory.fraction:
   Expressed in fraction, it is total storage that is reserved from JVM heap space.
5. spark.shuffle.file.buffer (Bytes):
   Buffer size (memory) for each output stream shuffle file. Expressed in KB.
6. spark.scheduler.listenerbus.eventqueue.capacity:
   This parameter determines the maximum number of events the event queue of the listener bus can handle.
7. spark.files.openCostInBytes:
   When multiple files are in the same partition, this is cost expressed as bytes to scan in the same time interval to open a file.
8. spark.storage.memoryMapThreshold(Bytes):
   Expressed in bytes, this parameter is the block size that Sparks memory maps when reading from a file.
9. Spark. files.maxPartitionBytes:
   When reading files, this parameter decides the size of the partition.
10. spark.default.parallelism:
    This parameter decides the number of partitions the RDD is to be divided into.
11. Spark.scheduler.mode:
    The mode in which to schedule submitted jobs.
    Following Table-II presents these parameters with their description and default value.

**Table II**

| Parameter Name | Default Value | Range |
|---|---|---|
| spark.task.cpus | 1 | [1,5] |
| spark.memory.storageFraction | 0.5 | [0.25,0.9] |
| spark.network.timeout (seconds) | 120 | [40,200] |
| spark.memory.fraction | 0.6 | [0.25,0.8] |
| spark.shuffle.file.buffer (Bytes) | 32k | [16k,512k] |
| spark.scheduler.listenerbus.eventqueue.capacity | 10000 | [2500,25000] |
| spark.files.openCostInBytes | 4194304 | [1048576, 16777216] |
| spark.storage.memoryMapThreshold(Bytes) | 2m | [1m, 5m] |
| spark.files.maxPartitionBytes | 134217728 | [33554432, 1073741824] |
| spark.default.parallelism | 24 | [4, 24] |
| spark.scheduler.mode | FIFO | [FIFO, FAIR] |

## IV. CLUSTER SETUP

***A. Hadoop:***

For Hadoop, the experiments were evaluated on a 3-node cluster (1 master node and 2 slave nodes). The master node has 8GB memory and Intel Core i7-3770 CPU @ 3.40GHz * 8 processor. The disk has 1TB of storage. Both the slave nodes have 4GB memory and Intel Core i7-3770 CPU @ 3.40GHz * 8 processor. The disk has 500 GB of storage.

***B. Spark:***

For Spark, experiments were evaluated on a 3-node cluster (1 master node and 2 slave nodes). Master node is equipped with Intel® Core™ i5-7200U CPU @ 2.50GHz × 4 processor, 8GB ram, GeForce 940MX/PCIe/SSE2 graphics card, 64-bit. Slave 1 is equipped with Intel® Core™ i5-4300U CPU @ 1.60Ghz processor, 4GB ram, Intel integrated graphics card, 64-bit. Slave 2 is equipped with Intel® Core™ i5-8300H CPU @ 2.30Ghz processor, 8GB ram, GeForce GTX 1050, 64-bit.

The experiments for both Hadoop and Spark were performed using the WordCount benchmark. The dataset used was 1GB.

## V. SYSTEM DESIGN

The entire code was divided into the following modules: 1. Configuration Manager and Performance Evaluator, 2. Grid Search with Finer Tuning, and 3. Controlled Random Search.

Configuration Manager and Performance Evaluator (CMPE) creates a level of abstraction between the algorithms, i.e., Grid Search with Finer Tuning and Controlled random search and the underlying technologies such as Hadoop and Spark. The CMPE module handles all the file handling, configuration changes, starting, stopping Hadoop and Spark, or other platform-specific operations. Grid Search with Finer Tuning module and Controlled Random Search implement the algorithm and pass the parameters and values to the CMPE module. The CMPE module further gets the time taken for the job to run and gives feedback to both the algorithm modules to be used further. A diagrammatic representation of these modules and the entire system design is shown in Figure I. The CMPE Module consists of both Job Runner and Performance Evaluator, as shown in Figure I. The rest two modules are explicitly shown. The Admin actions for Tuning the parameters by selecting the algorithm to be run and the platform to run. This then triggers the algorithm on the chosen platform, and after the successful run of the algorithm, the best configuration obtained is returned. The CMPE module also has a log file provision that writes each run's results in the file with the parameter set and the time required for the run.

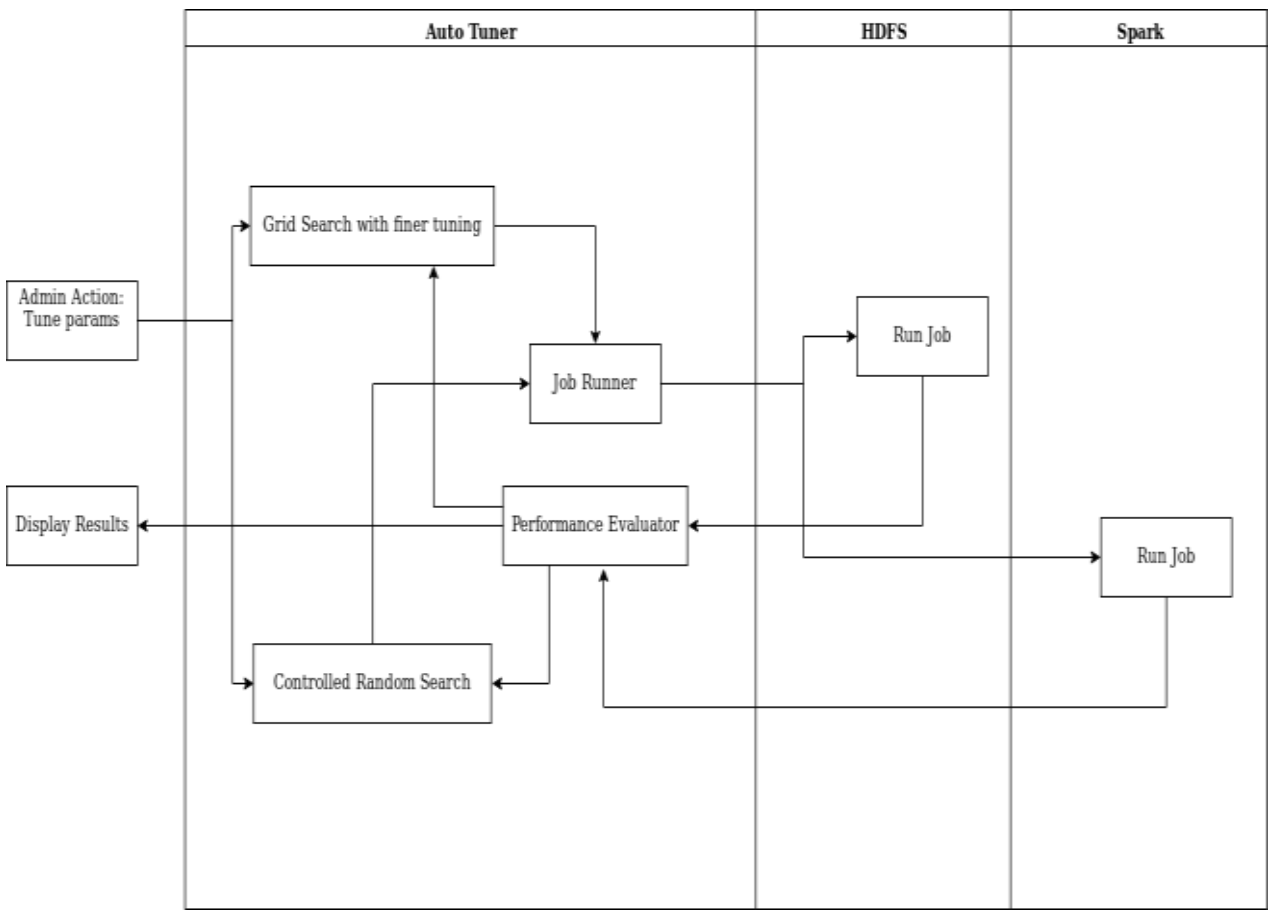

**FIGURE I**

## VI. INITIAL EXPERIMENT

### *A. Hadoop*

All the shortlisted parameters were kept at their default value, and then the job was run. The results obtained are as shown below in Table III.

**Table III**

| Parameter Name | Tuned Value | Execution Time |
|---|---|---|
| mapreduce.map.memory.mb | 1024 | 5 min 39 sec |
| dfs.blocksize | 128 | |
| mapreduce.tasktracker.map.tasks.maximum | 2 | |
| mapreduce.job.reduce.slowstart.completedmaps | 0.05 | |
| mapreduce.map.output.compress | FALSE | |
| mapreduce.job.reduces | 1 | |
| mapreduce.task.io.sort.mb | 100 | |
| mapreduce.job.maps | 2 | |
| mapreduce.task.io.sort.factor | 10 | |
| dfs.replication | 3 | |
| mapreduce.tasktracker.reduce.tasks.maximum | 2 | |
| mapreduce.job.jvm.numtasks | 1 | |

Another experiment kept one parameter at its optimal value (value obtained from previous research work of manual tuning) and the rest all at default. The results obtained are shown in Table IV below.

**Table IV**

| Parameter Name | Tuned Value | Execution Time | Improvement (%) |
|---|---|---|---|
| mapreduce.map.memory.mb | 3072 | 3 min 39 sec | 35.4% |
| dfs.blocksize | 256 | 4 min 33 sec | 19.47 % |
| mapreduce. Task tracker.map. tasks.maximum | 64 | 4 min 36 sec | 18.58 % |
| mapreduce.job.reduce.slowstart.completed maps | 0.9 | 19 min 34 sec | - |
| mapreduce.map.output.compress | TRUE | 3 min 38 sec | 35.69 % |
| mapreduce.job.reduces | 1 | 4 min 34 sec | 19.17 % |
| mapreduce.task.io.sort.mb | 32 | 4 min 21 sec | 23.01 % |
| mapreduce.job.maps | 32 | 4 min 34 sec | 19.17 % |
| mapreduce.task.io.sort.factor | 10 | 4 min 53 sec | 13.57 % |
| dfs.replication | 1 | 4 min 23 sec | 22.42 % |
| mapreduce.tasktracker.reduce.tasks.maximum | 16 | 4 min 42 sec | 16.81 % |
| mapreduce.job.jvm.numtasks | 16 | 4 min 36 sec | 18.58 % |

The next experiment performed was by keeping all the parameters at their optimal values (value obtained from previous research work of manual tuning). The results are as shown below in Table V.

**Table V**

| Parameter Name | Tuned Value | Execution Time | Improvement (%) |
|---|---|---|---|
| mapreduce.map.memory.mb<br>dfs.blocksize<br>mapreduce.tasktracker.map.tasks.maximum<br>mapreduce.job.reduce.slowstart.completedmaps<br>mapreduce.map.output.compress<br>mapreduce.job.reduces<br>mapreduce.task.io.sort.mb<br>mapreduce.job.maps<br>mapreduce.task.io.sort.factor<br>dfs.replication<br>mapreduce.tasktracker.reduce.tasks.maximum<br>mapreduce.job.jvm.numtasks | 3072<br>256<br>64<br>0.9<br>TRUE<br>1<br>32<br>32<br>10<br>1<br>16<br>16 | 2 min 15 sec | 60.18 % |

***B. Spark***

All the shortlisted parameters were kept at their default value, and then the job was run. The results obtained are as shown below in Table VI.

**Table VI**

| Parameter Name | Tuned Value | Execution Time |
|---|---|---|
| spark.task.cpus<br>spark.memory.storageFraction<br>spark.network.timeout (sec)<br>spark.memory.fraction<br>spark.shuffle.file.buffer (Bytes)<br>spark.scheduler.listenerbus.eventqueue.capacity<br>spark.files.openCostInBytes<br>spark.storage.memoryMapThreshold (Bytes)<br>spark.files.maxPartitionBytes<br>spark.default.parallelism<br>spark.scheduler.mode | 1<br>0.5<br>120<br>0.6<br>32k<br>10000<br>4194304<br>2m<br>134217728<br>24<br>FIFO | 1 min 57 sec |

Another experiment kept one parameter at its optimal value (value obtained from previous research work of manual tuning) and the rest all at default. The results obtained are shown in Table VII below.

**Table VII**

| Parameter Name | Tuned Value | Execution Time | Improvement (%) |
|---|---|---|---|
| spark.task.cpus | 5 | 29 sec | 75.21% |
| spark.memory.storageFraction | 0.6 | 1 min 49 sec | 6.83% |
| spark.network.timeout(sec) | 160 | 1 min 51 sec | 5.12% |
| spark.memory.fraction | 0.25 | 1 min 52 sec | 4.27% |
| Spark.shuffle.file.buffer (Bytes) | 256k | 1 min 52 sec | 4.27% |
| spark.scheduler.listenerbus.eventqueue.capacity | 10000 | 1 min 57 sec | 0% |
| spark.files.openCostInBytes | 1048576 | 1 min 56 sec | 0.8% |
| spark.storage.memoryMapThreshold (Bytes) | 1m | 1 min 59 sec | - |
| spark.files.maxPartitionBytes | 33554432 | 2 min 1 sec | - |
| spark.default.parallelism | 12 | 1 min 55 sec | 1.7% |
| spark.scheduler.mode | FAIR | 1 min 56 sec | 0.8% |

The next experiment performed was by keeping all the parameters at their optimal values (value obtained from previous research work of manual tuning). The results are shown below in Table VIII.

**Table VIII**

| Parameter Name | Tuned Value | Execution Time |
|---|---|---|
| spark.task.cpus | 5 | 27 sec |
| spark.memory.storageFraction | 0.6 | |
| spark.network.timeout (sec) | 160 | |
| spark.memory.fraction | 0.25 | |
| spark.scheduler.listenerbus.eventqueue.capacity | 256k | |
| spark.files.openCostInBytes | 10000 | |
| spark.storage.memoryMapThreshold (Bytes) | 1048576 | |
| spark.files.maxPartitionBytes | 1m | |
| spark.default.parallelism | 3355443212 | |
| spark.scheduler.mode | FAIR | |

## VII. CMPE MODULE

Configuration Manager and Performance Evaluator, as known as CMPE, is a helper module between algorithm and system. It helps the algorithms to interact with the systems via CMPE. CMPE manages the configuration and evaluates the results, and gives us the optimal configuration of the parameters. By setting these values of the parameters, a reduction in execution time of the job is observed, hence improving the system's performance on the other hand. CMPE first identifies the platform (Hadoop/Spark), and then it changes the configuration file by identifying which parameter lies in which file for the Hadoop. And for Spark, it runs the commands on the terminal. Before running the Hadoop, the safe mode needs to be disabled because it doesn't allow the user to make changes in the configuration file when on safe mode. The output directory is also deleted because having two same-named output directories generates an error. CMPE also maintains a log file. Logfile keeps track of which function is running at what time. It also keeps the result of every run of the algorithm. Analyzing the log file helps in finding the optimal configuration. System time is taken at appropriate places to calculate the total run time or maintain a log file.

## VIII. ALGORITHM - I

Grid Search is also known as parameter sweep, and this algorithm means an exhaustive search on the entire parameter space. "A grid search algorithm must be guided by some performance metric, typically measured by cross-validation on the training set [3] or evaluation on a held-out validation set."[4]

"Since the parameter space of a machine learner may include real-valued or unbounded value spaces for certain parameters, manually set bounds and discretization may be necessary before applying grid search."

Grid search suffers from the curse of dimensionality but is often embarrassingly parallel because the hyperparameter settings it evaluates are typically independent of each other.[5]

One of the major debates when Grid search is brought into the picture, is this algorithm's problem due to equally separated sampling. A lot of research has been done on Grid Search in various fields and various applications. However, this paper implements a flavor of Grid Search called Grid Search with Finer Tuning to overcome this problem.

Grid Search with Finer Tuning is essentially a Grid Search with a Finer Tuning overhead. Finer Tuning makes sure the probability of Grid search due to even sampling and missing out an optimal configuration lying in between the two sampled values is reduced.

In this case, the parameters are of two types: Continuous and Boolean. For all continuous parameters, the algorithm first samples the data with a predefined 'step' value. For the Boolean parameters, no change is done. After sampling it evenly, a parameter grid list is created from these parameters. Grid Search is then applied to this parameter space to find out the best configuration. Now for finer tuning, let's introduce a concept of 'most influential parameters. These are the parameters that have a huge effect on the performance of Hadoop as well as Spark and are found based on the previous work as mentioned in section II. The most influential parameters found were - mapreduce.map.memory.mb and dfs.blocksize for Hadoop and spark.memory.storageFraction and spark.network.timeout for Spark.

Finer tuning is applied to these parameters. After finding out the best configuration via Grid Search, a finer tuning is applied across these 'most influential parameters, and the algorithm is run again. In finer tuning, a finer sampling is done around the best configuration obtained in the previous run. Finer tuning is carried out by sampling the data again for the most influential parameters, which is done by calculating:

new_lower_bound= best_value- (old_lower_bound / 2)
new_upper_bound= best_value+ (old_lower_bound/ 2)
increment = new_lower_bound/ 2

The time complexity of this algorithm is: O(nm+k)
where n is the number of parameters,
m is the number of steps,
k is the number of parameters for finer tuning.

The algorithm is as shown below.

Algorithm: Grid Search with Finer Tuning

```
Generate param_grid of nelements of the applicable values of parameters
Setparam_grid_list= getParameterGridList(param_grid)
for i = 0 → m do
        Tune the Configurations of Hadoop/Spark as per
                param_grid_list[i]
        Run the job and calculate execution_time
        if execution_time<min_execution_timedo
                min_execution_time= execution_time
                best_config= param_grid_list[i]
        end if
end for
for i = 0 → l do
        param = most_influential_params[l]
        lower_bound= param_grid[param][0]
        num = sizeOf(param_grid[param])/sizeOf(int)
        upper_bound= param_grid[param][num]
        new_param_grid[param] = sampleData(lower_bound,
        upper_bound)
end for
for param in best_configdo
        if param not in most_influential_paramsdo
                new_param_grid[param] =
        best_config[param]
        end if
end for
new_param_grid_list= getParamGridList(new_param_grid)
for i = 0 → sizeOf(new_param_grid_list)do
        Tune the Configurations of Hadoop/Spark as per
        new_param_grid_list[i]
        Run the job and calculateexecution_time
        if execution_time<min_execution_timedo
                min_execution_time= execution_time
                best_config= new_param_grid_list[i]
        end if
end for
return best_config
```

### *A. Hadoop*

Taking into consideration the total number of parameters shortlisted for Hadoop, i.e., 12, and sampling of each parameter for 10 different values at least, the total number of configuration space would have reached the power of $10^{12}$. Considering the default configuration time that was 5 min 39 secs, it was obviously not feasible. Hence taking into consideration minimal parameters due to the tradeoff between the total time taken for the run and the optimal configuration, the closest to optimal was obtained by using the following parameters - mapreduce.map.memory.mb, dfs.blocksize, mapreduce.task tracker.map.tasks.maximum, mapreduce.map.output. Compress, dfs. Replication.

From the previous work and initial experiments, it was found that dfs. Replication gave its optimum when the value was set to 1 and mapreduce.map.output. Compress gave its optimum when the value was set to TRUE. Hence these values were set throughout the run to the mentioned above.

According to the experimentation as well as previous works, below is a list of the most influential parameters of Hadoop that are used for finer tuning in this Grid search algorithm.

1. mapreduce.map.memory.mb:

   This parameter is varied with a step of 32 during finer tuning.

2. dfs.blocksize

   This parameter is varied with a step of 8 during finer tuning.

Running the algorithm on the cluster gave the optimal configuration, as shown in Table IX.

**Table IX**

| Parameter Name | Tuned Value | Time taken | % Optimization |
|---|---|---|---|
| mapreduce.map.memory.mb | 512 | 1 min 39 sec | 70.8% |
| dfs.blocksize(Bytes) | 192 | | |
| mapreduce.tasktracker.map.tasks.maximum | 128 | | |
| mapreduce.job.reduce.slowstart.completedmaps | 0.05 | | |
| mapreduce.map.output.compress | TRUE | | |
| mapreduce. Job. reduces | 1 | | |
| mapreduce.task.io.sort.mb | 100 | | |
| mapreduce. Job. maps | 2 | | |
| mapreduce.task.io.sort.factor | 10 | | |
| dfs.replication | 1 | | |
| mapreduce.tasktracker.reduce.tasks.maximum | 2 | | |
| mapreduce.job.jvm.numtasks | 1 | | |

### *B. Spark*

According to the shortlisted parameters for Spark, sampling of each parameter is done for the parameters: spark. Task. cpus, spark. memory.storageFraction, spark. network.timeout, spark.shuffle.file.buffer, spark. Memory. Fraction, while the parameter spark.scheduler.mode is set to FAIR since this type of scheduler mode, is found to be much more promising than the FIFO mode.

For finer tuning, the most influential parameters, i.espark.memory.storageFraction, and spark.network.timeout, is considered as the parameter spark. Task. cpus gives the best outcome at its max value; hence varying it will only lead to a decrease in performance.

According to the experimentation as well as previous works, below is a list of the most influential parameters of Hadoop that are used for finer tuning in this Grid search algorithm.

1. Spark.task.cpus

   This parameter, though it is most influential, gives the best performance when it is set maximum. This directly influences parallel computing. Hence during the finer tuning, this parameter is always set at the maximum and is not varied.

2. spark.memory.storageFraction

   This parameter is varied with a step of 0.25 during finer tuning.

3. Spark.network.timeout

   This parameter is varied with a step of 20 during finer tuning.

Running the algorithm on the cluster gave the optimal configuration, as shown in Table X.

**Table X**

| Parameter Name | Tuned Value | Execution Time | Improvement Percentage |
|---|---|---|---|
| spark.task.cpus | 5 | 22secs | 81.19% |
| spark.memory.storageFraction | 0.9 | | |
| spark.network.timeout (sec) | 120 | | |
| spark.memory.fraction | 0.5 | | |
| spark.shuffle.file.buffer (Bytes) | 128k | | |
| spark.scheduler.listenerbus. eventqueue.capacity | 10000 | | |
| spark.files.openCostInBytes | 4194304 | | |
| spark.storage. memoryMapThreshold (Bytes) | 2m | | |
| spark.files.maxPartitionBytes | 134217728 | | |
| spark.default.parallelism | 24 | | |
| spark.scheduler.mode | FAIR | | |

## IX. ALGORITHM – II

Random Search is an algorithm that does not depend on derivatives. Rather it is a direct search in a continuous domain. "It can outperform Grid search, especially when only a small number of hyperparameters affects the final performance of the machine learning algorithm.[5] In this case, the optimization problem is said to have a low intrinsic dimensionality".[7][4]

In Random Search, the main objective of finding the global minimum or maximum is achieved by sampling the search space randomly. Random Search is able to return a reasonable approximation of the optimal solution, but as the dimension of the search space increases, it requires more time. A flavor of Random Search called Controlled

Random Search that gives random search more power and accuracy to reach towards the objective is used here.
Controlled Random Search combines random search and mode-seeking routines into one process. The Controlled Random Search algorithm runs the simple random search and obtains new bounds for the further run. It keeps running the process of random search with new bounds of parameters till it reaches a threshold.[8]
As already mentioned, the parameter values are of two types: continuous and Boolean. In this algorithm, the random values are selected from the continuous domain for each parameter. For Boolean values, randomly, either TRUE or FALSE is chosen. By taking these randomly chosen values considering the lower and upper bound of each parameter, a configuration list is formed. These configuration lists are used to tune the frameworks from which some specified number of configurations lists are selected using criteria of execution time. From these selected configurations minimum and maximum of each parameter are found out and are called the new upper and lower bound.

Upper_bound[param] = max(configuration_lists)
Lower_bound[param] = min(configuration_lists)

These newly formed bounds are again fed to the algorithm to form new configuration lists. This is done until a threshold is not broken. The threshold is set to minimum execution time from the configuration list before calculating the new upper and lower bound.
The time complexity of this algorithm is: O(nm)
where n is the number of random values generated,
m is the value for reaching a threshold.

The algorithm is as shown below.

Algorithm: Controlled Random Search

```
Function random_search(lower_bound, upper_bound)
        Set param_grid_list =
        getRandomizedParameterGridList(lower_bound,
        upper_bound)
        Set num = 0
        for i = 0 → m do
          Tune the Configurations of Hadoop/Spark as per
        param_grid_list[i]
          Run the job and calculate execution_time
          if execution_time<min_execution_time do
                min_execution_time = execution_time
                 best_config[num] = param_grid_list[i]
                 num = num + 1
          end if
        end for
        Select top k best_config and remove the rest
        return best_config

Function Controlled_randomSearch()
        Set best_config = None
        revised_best_config = random_search(lower_bound,
        upper_bound)
        while variation(revised_best_config, best_config) >
        threshold do
        Set m =  number of most_influential_params
        for i = 0 → k do
                 for j = 0 → m do
        lower_bound[param] =  min(best_config[param])
        upper_bound[param] =  max(best_config[param])
        end for
        end for
        best_config = revised_best_config
        revised_best_config = random_search(lower_bound,
        upper_bound)
        end while
        return best_config
```

### *A. Hadoop*

Considering the selected parameters, its default upper and lower bound were fed to the algorithm initially.

**Table XI**

| Parameter Name | Tuned Value | Time Taken | % Optimization |
|---|---|---|---|
| mapreduce.map.memory.mb | 1736 | 2mins 51 sec | 49.55% |
| dfs.blocksize(Bytes) | 192 | | |
| mapreduce.tasktracker.map.tasks.maximum | 128 | | |
| mapreduce.job.reduce.slowstart.completedmaps | 0.05 | | |
| mapreduce.map.output.compress | TRUE | | |
| mapreduce. Job. reduces | 1 | | |
| mapreduce.task.io.sort.mb | 59 | | |
| mapreduce. Job. maps | 20 | | |
| mapreduce.task.io.sort.factor | 26 | | |
| dfs.replication | 1 | | |
| mapreduce.tasktracker.reduce.tasks.maximum | 21 | | |
| mapreduce.job.jvm.numtasks | 1 | | |

From the previous run, dfs. Replication and dfs. Blocksize, when set to 1and 192, gave optimum performance. So, to decrease the load on running this algorithm, these two parameters were set to the above-mentioned value.

Table XI shows the results of Controlled Random Search on Hadoop.

### *B. Spark*

Considering the upper and lower bound of the selected spark parameters, random values of them were generated and fed to the algorithm. The spark.scheduler.mode takes two types of values: FAIR and FIFO, so the code chooses one of them randomly.

**Table XII**

| Parameter Name | Tuned Value | Execution Time | Improvement (%) |
|---|---|---|---|
| spark.task.cpus | 5 | 26 sec | 77.77 % |
| spark.memory.storageFraction | 0.82 | | |
| spark.network.timeout (sec) | 137 | | |
| spark.memory.fraction | 0.77 | | |
| spark.shuffle.file.buffer (Bytes) | 128k | | |
| spark.scheduler.listenerbus.eventqueue.capacity | 16343 | | |
| spark.files.openCostInBytes | 11213009 | | |
| spark.storage.memoryMapThreshold (Bytes) | 2m | | |
| spark.files.maxPartitionBytes | 36300419 | | |
| spark.default.parallelism | 14 | | |
| spark.scheduler.mode | FAIR | | |

## X. Analysis of Results

Most Influential Parameters:

In the case of both Hadoop and Spark, some parameters have much more impact on the performance than other parameters. This comparison, however, is very important to determine the finer tuning required in Hadoop as well as spark. According to the experimentation as well as previous works, below is a list of the most influential parameters that are used for finer tuning in this Grid search algorithm.

Hadoop:

• mapreduce.map.memory.mb

This parameter is varied with a step of 32 during finer tuning.

• dfs.blocksize

This parameter is varied with a step of 8 during finer tuning.

Spark:

• spark.task.cpus

This parameter, though it is most influential, gives the best performance when it is set maximum. This directly influences parallel computing. Hence during the finer tuning, this parameter is always set at the maximum and is not varied.

• spark.memory.storageFraction

This parameter is varied with a step of 0.25 during finer tuning.

• spark. Network. timeout

This parameter is varied with a step of 20 during finer tuning.

Statistical Representation of Results:

### *A. Hadoop*

#### *a) Default vs. All at Individual Optimal*

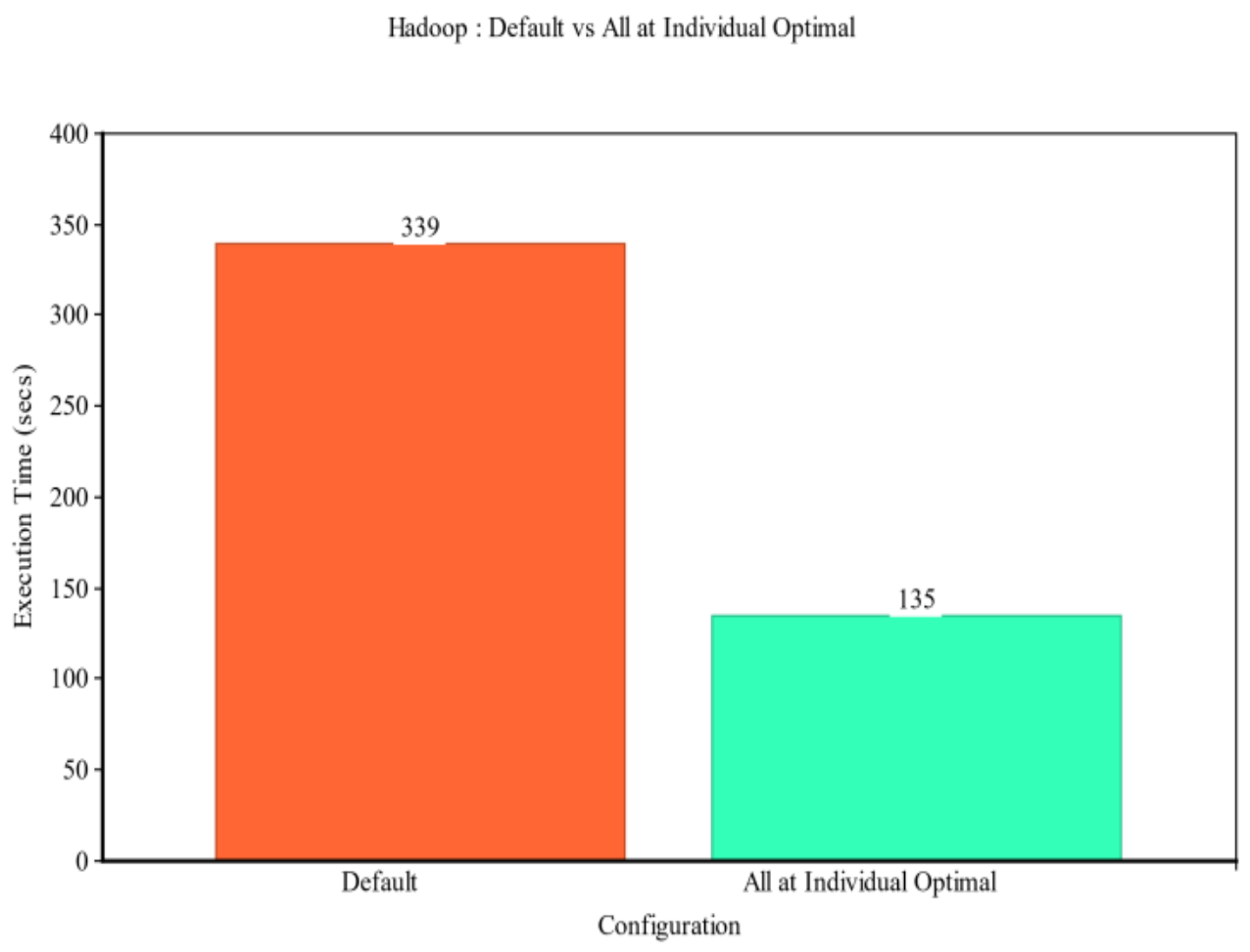

**Figure 1: Hadoop: Default vs. All at Individual Optimal**

#### *b) One Optimal and others at Default*

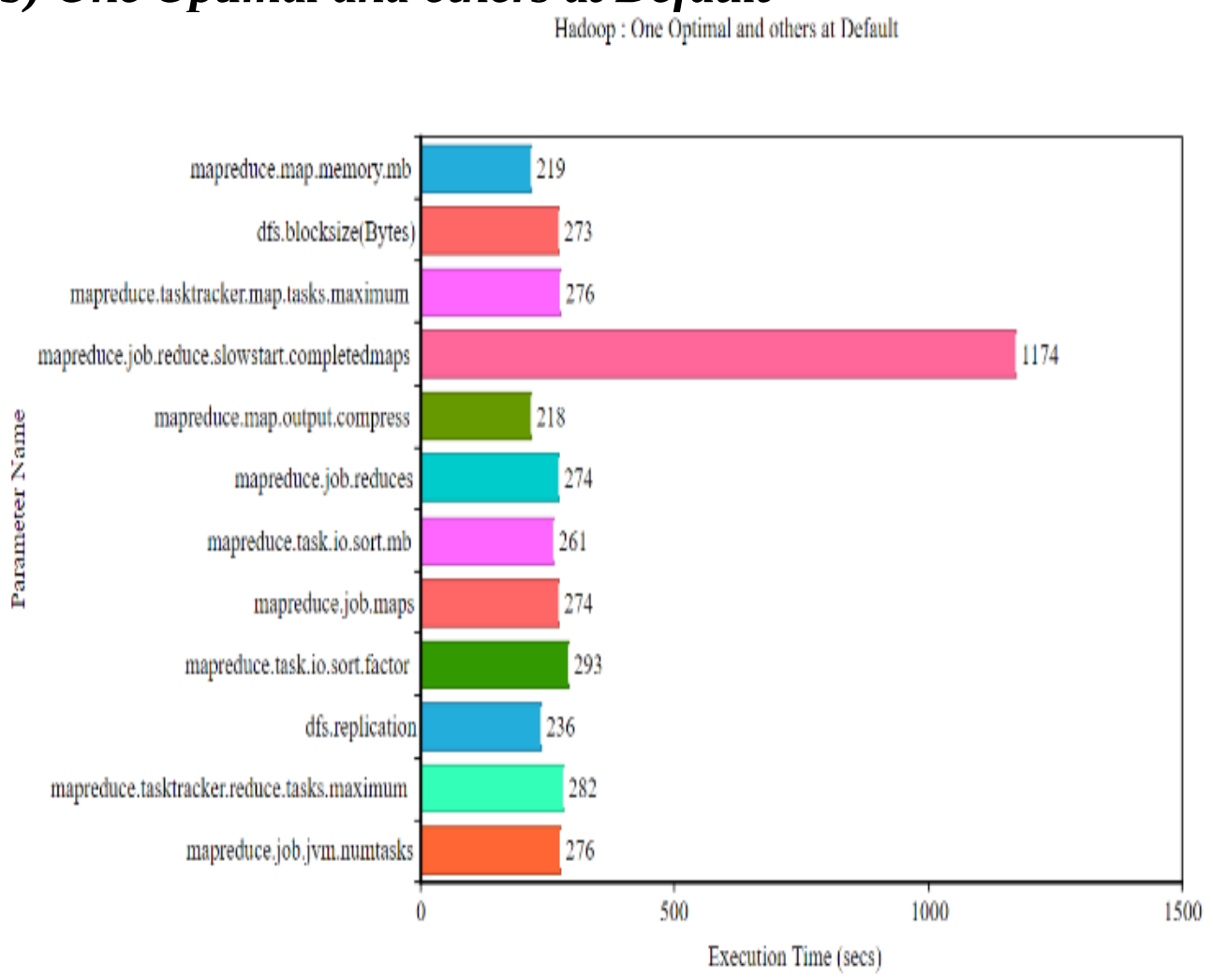

**Figure 2: Hadoop: One Optimal and others at Default**

### *c) Time Required Vs. Configurations*

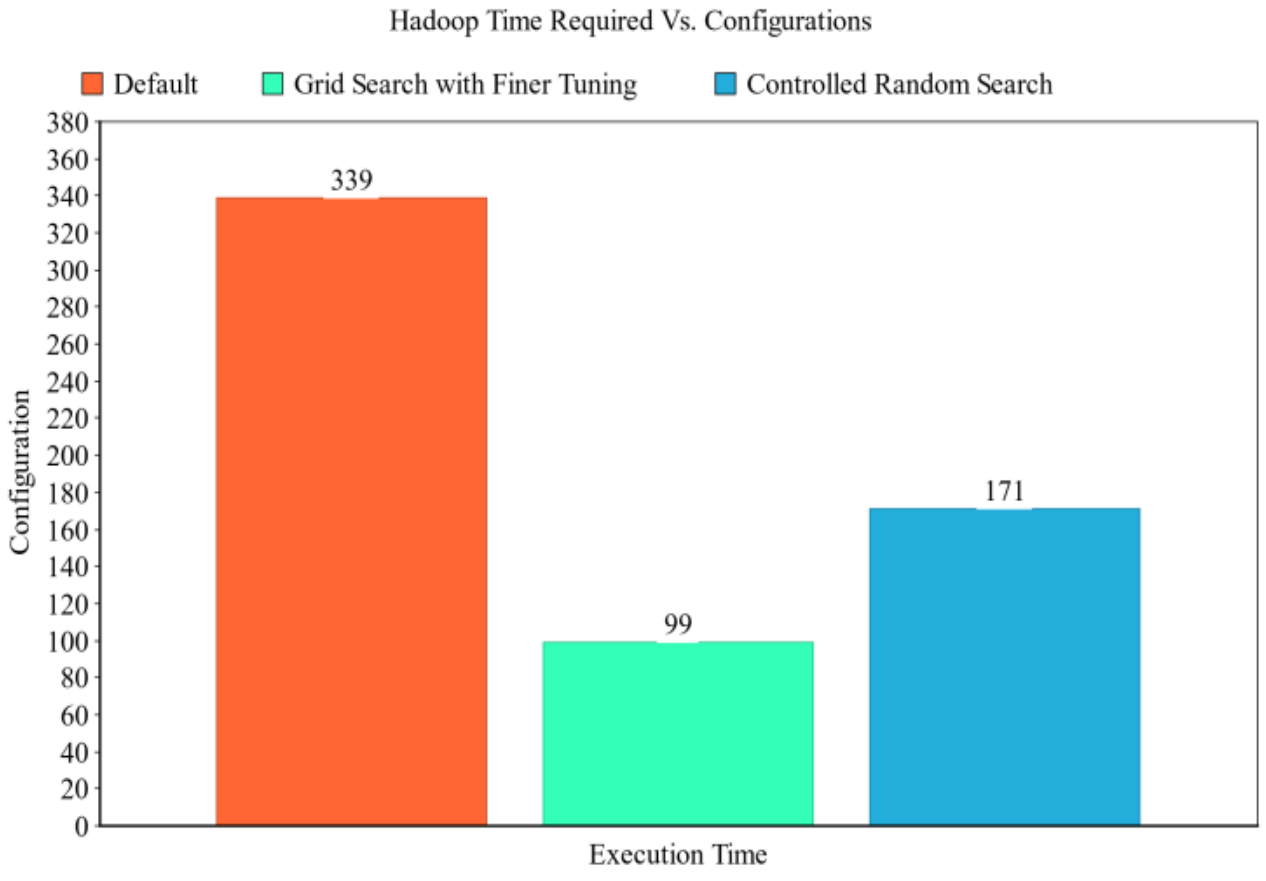

**Figure 3: Hadoop: Time Required Vs. Configurations**

### *B. Spark*

### *a) Default vs. All at Individual Optimal*

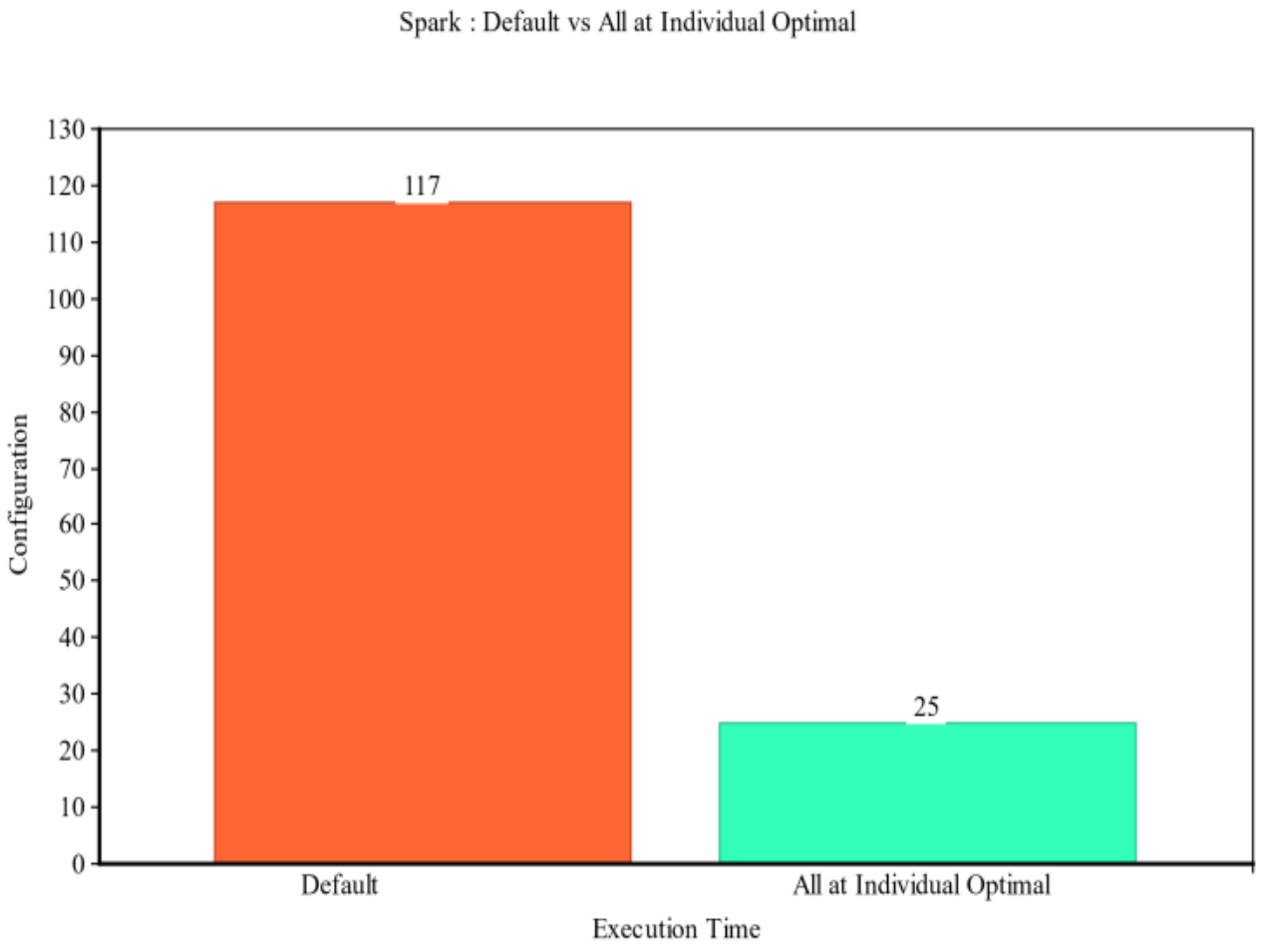

**Figure 4: Spark: Default vs. All at Individual Optimal**

### *b) One Optimal and others at Default*

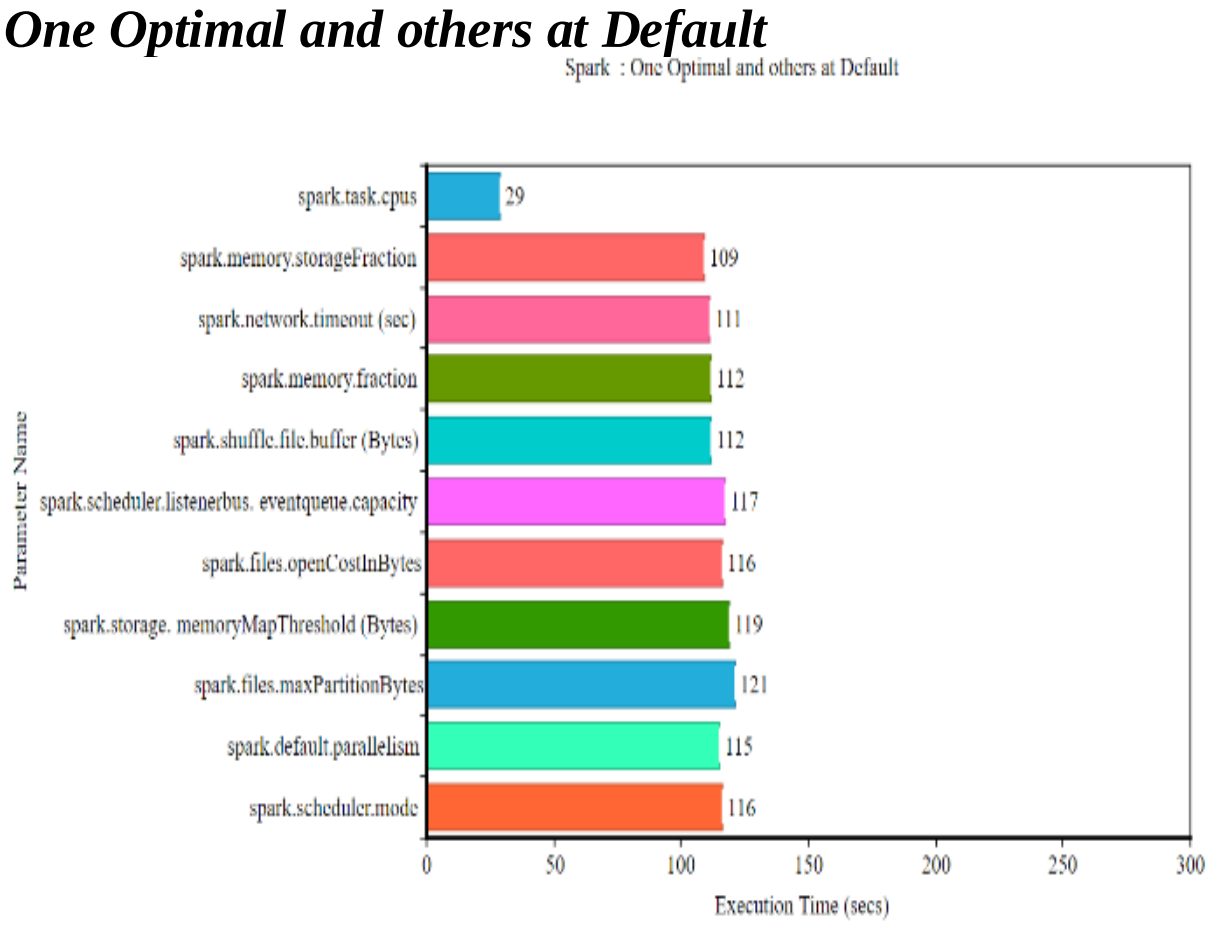

**Figure 5: Spark: One Optimal and others at Default**

### *c) Time Required Vs. Configurations*

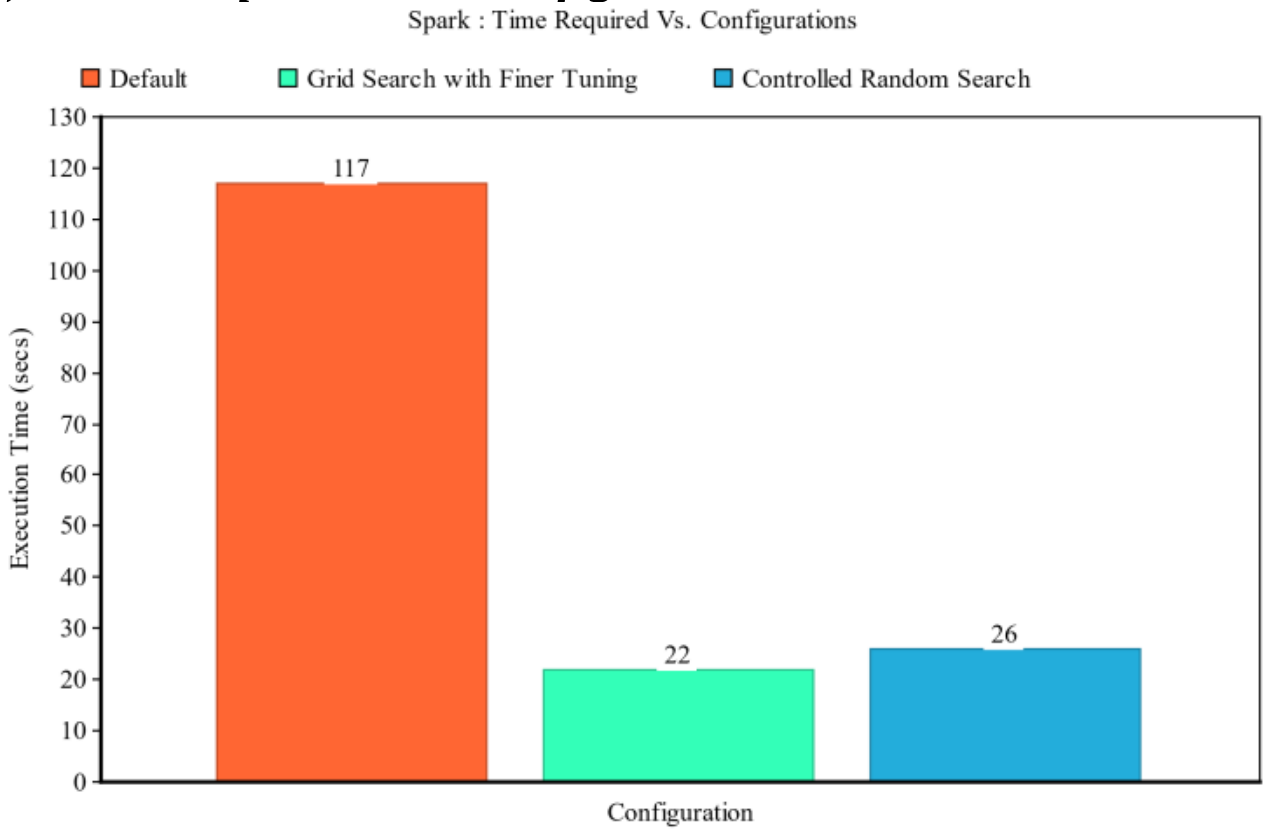

**Figure 6: Spark: Time Required Vs. Configurations**

### *C. Comparison*

### *a) Default Configuration of Hadoop and Spark*

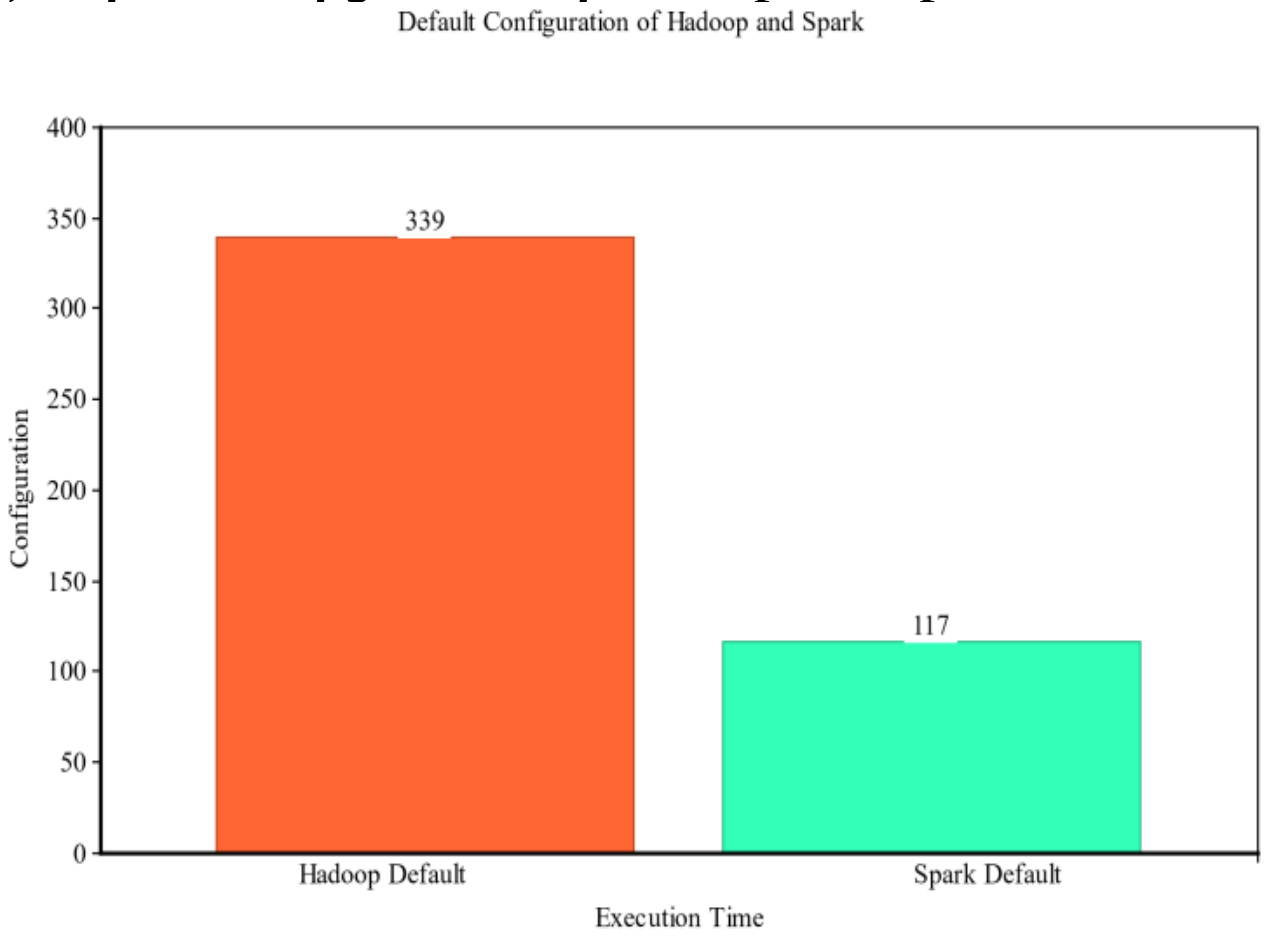

**Figure 7: Default Configuration of Hadoop and Spark**

### *b) Comparison of two Algorithms(Hadoop)*

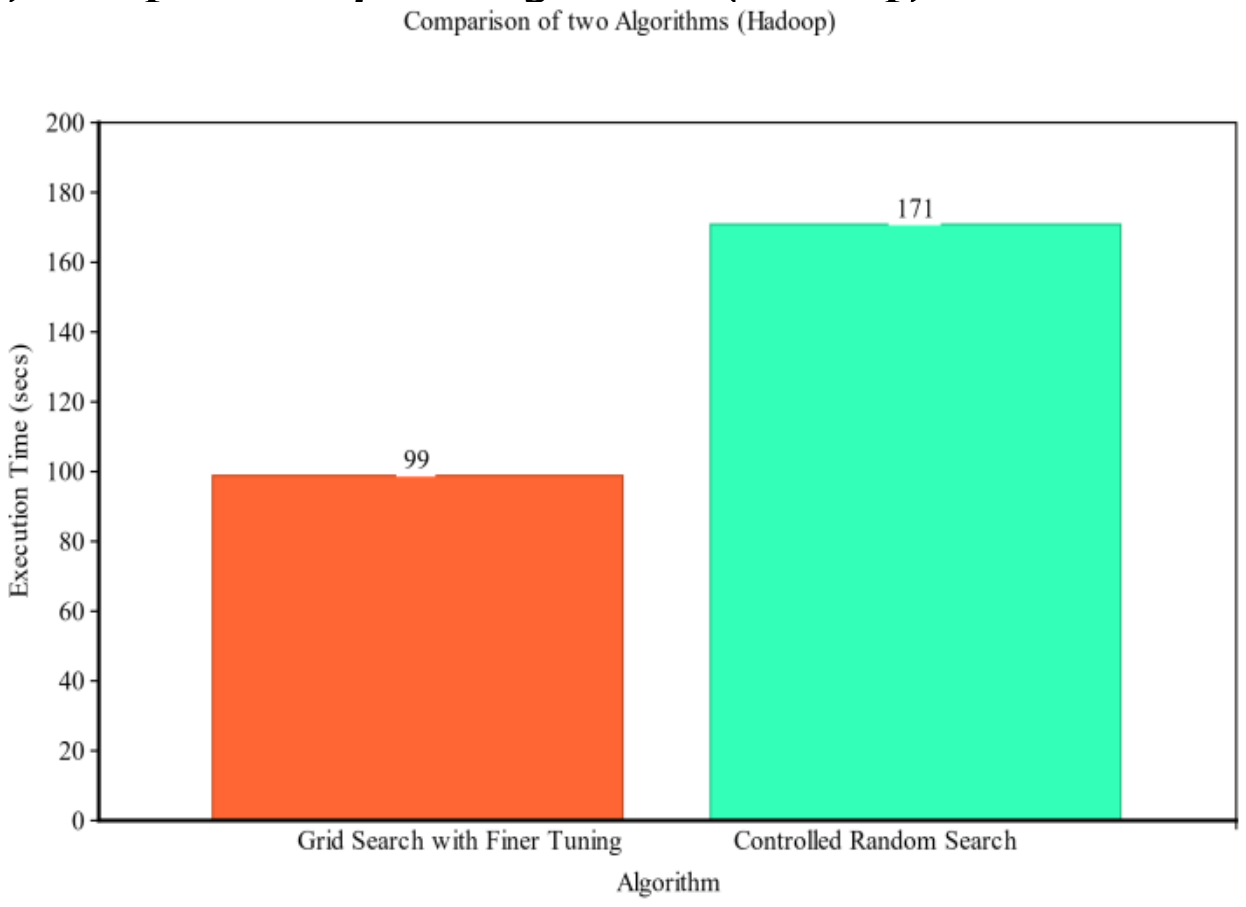

**Figure 8: Comparison of two Algorithms (Hadoop)**

### c) Comparison of two Algorithms (Spark)

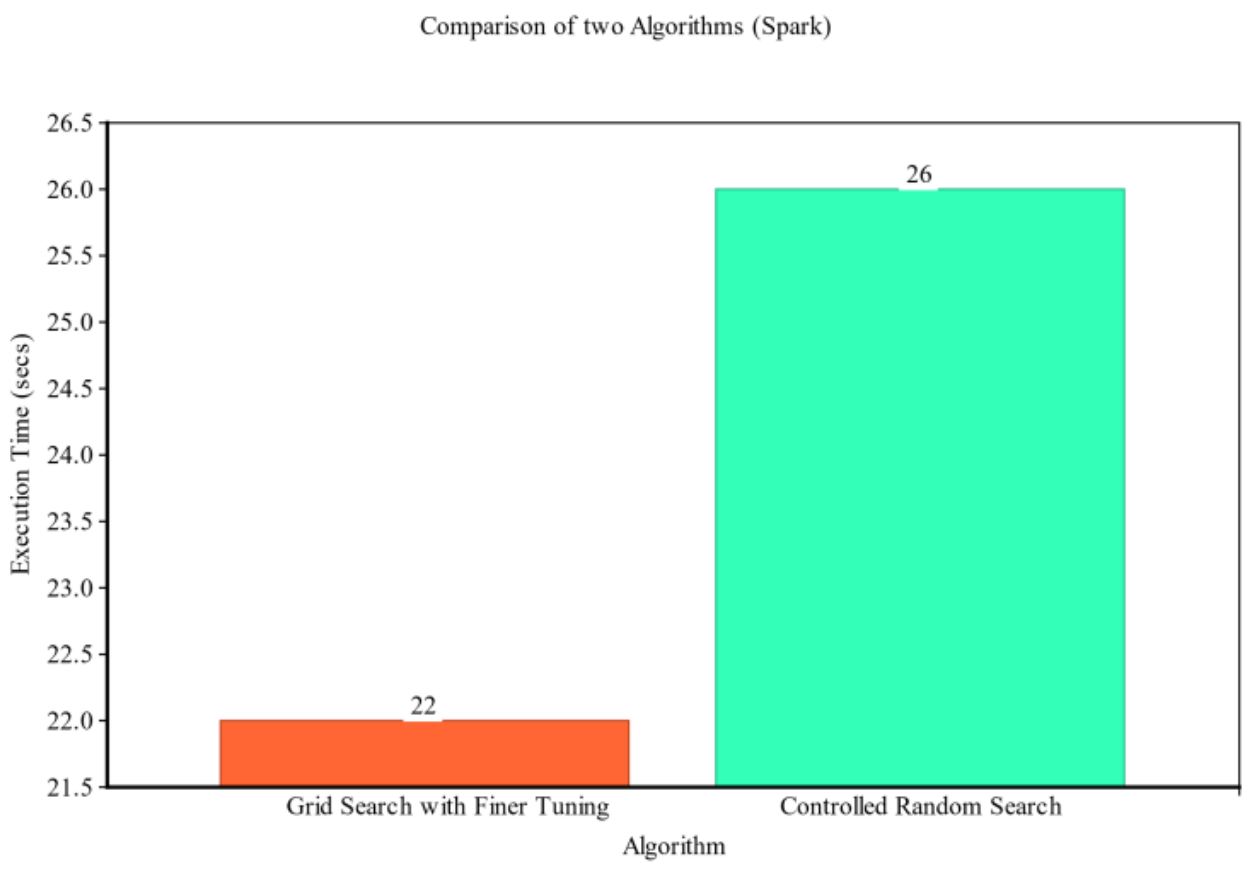

**Figure 9: Comparison of two Algorithms (Spark)**

## XI. COMPARISON

| Parameters of comparison | Grid Search with Finer Tuning | Controlled Random Search |
|---|---|---|
| Reduction in execution time (Hadoop) | 70.8% | 49.55% |
| Reduction in execution time (Spark) | 81.19% | 77.77% |

## XII. CONCLUSION

Auto-tuning of parameters has become as obligatory as finding the optimal configuration to improve the performance of the system. This paper, after analyzing the importance of the automatic tuning of parameters, proposes two algorithms: Grid Search with Finer Tuning and Controlled Random Search. These algorithms have made use of the preceding work, which helped in selecting those parameters which influenced the overall performance of the system. The results obtained demonstrated that Grid Search with Finer Tuning suits well for spark as compared to Hadoop. Grid Search with Finer Tuning has shown great speed for Spark. Controlled Random Search, however not much useful to Hadoop, has shown considerable improvements in Spark. Also, in both cases, Grid search with finer tuning has proved to be much more efficient. Finally, both the algorithms proposed in this paper are capable of finding efficient configurations and thus reducing the execution time required while running a single job on Hadoop and Spark platforms.

## XII. FUTURE SCOPE

The entire research is focused on optimizing the performance of Hadoop and Spark with respect to time. This paper does not take into consideration other aspects of optimization such as optimization with respect to resources, cpu, memory, or cost. A lot of work can be done considering these aspects as well.
The algorithms used have their own benefits and drawbacks, as mentioned above. A better algorithm can be drafted and implemented that can combine the plus points of both these algorithms and eliminate the drawbacks. Future work could include working on such algorithms and considering a multi-objective approach.

## XIV. REFERENCES

[1] https://hadoop.apache.org/
[2] https://spark.apache.org/
[3] Chin-Wei Hsu, Chih-Chung Chang and Chih-Jen Lin (2010). A practical guide to support vector classification. Technical Report, National Taiwan University.
[4] https://en.wikipedia.org/wiki/Hyperparameter_optimization
[5] Bergstra, James, and YoshuaBengio., Random search for hyper-parameter optimization., The Journal of Machine Learning Research 13(1) (2012) 281-305.
[6] Armanur Rahman, J.Hossen, Venkataseshaiah C., SMBSP: A Self-Tuning Approach using Machine Learning to Improve Performance of Spark in Big Data Processing., 7th International Conference on Computer and Communication Engineering (ICCCE) (2018).
[7] Wang, Ziyu, et al., Bayesian optimization in a billion dimensions via random embeddings., Journal of Artificial Intelligence Research 55 (2016) 361-387.
[8] W.L. Price., A Controlled Random Search Procedure for global optimization.
[9] Yigitbasi, Nezih, et al., Towards machine learning-based auto-tuning of mapreduce. 2013 IEEE 21st International Symposium on Modelling, Analysis, and Simulation of Computer and Telecommunication Systems. IEEE, (2013).
[10] Ding, Xiaoan, Yi Liu, and Depei Qian., Jellyfish: Online performance tuning with adaptive configuration and elastic container in Hadoop yarn., In 2015 IEEE 21st International Conference on Parallel and Distributed Systems (ICPADS), IEEE, (2015) 831-836.
[11] Bansal, Garvit, Anshul Gupta, Utkarsh Pyne, Manish Singhal, and Subhasis Banerjee., A framework for performance analysis and tuning in Hadoop-based clusters., In Smarter Planet and Big Data Analytics Workshop (SPBDA 2014), held in conjunction with International Conference on Distributed Computing and Networking (ICDCN 2014), Coimbatore, INDIA. (2014).
[12] Gu, Jing, Ying Li, Hongyan Tang, and Zhonghai Wu., Auto-Tuning Spark Configurations Based on Neural Network., In 2018 IEEE International Conference on Communications (ICC), (2018) 1-6. IEEE.
[13] Du, Haizhou, Ping Han, Wei Chen, Yi Wang, and Chenlu Zhang., Otterman: A Novel Approach of Spark Auto-tuning by a Hybrid Strategy., In 2018 5th International Conference on Systems and Informatics (ICSAI), IEEE, (2018) 478-483.
[14] Nguyen, Nhan, Mohammad Maifi Hasan Khan, and Kewen Wang., Towards automatic tuning of apache spark configuration., 2018 IEEE 11th International Conference on Cloud Computing (CLOUD). IEEE, (2018).
[15] Shivarkar, Sayali Ashok,. Speed-up extension to Hadoop system., International Journal of Engineering Trends and Technology (IJETT) (2014).